\documentclass[a4paper,fleqn,usenatbib]{mnras}
\usepackage[T1]{fontenc}
\usepackage{ae,aecompl}
\usepackage{epsfig,graphicx}	
\usepackage{amsmath}	
\usepackage{amssymb}	
\usepackage{multicol}
\usepackage{natbib}

\title[Optical analysis of the CMB CSc-1]{Optical analysis of a CMB cosmic string candidate}

\author[O. S.~Sazhina et al.]{O. S.~Sazhina,$^{1}$\thanks{E-mail: cosmologia@yandex.ru}
D. Scognamiglio,$^{2}$
M.V. Sazhin,$^{1}$
M. Capaccioli$^{2}$\\
\\
$^{1}$Sternberg Astronomical Institute of Lomonosov Moscow State University, Universitetskiy pr. 13, 119234 - Moscow, Russia \\
$^{2}$Dipartimento di Scienze Fisiche, Universit\`a di Napoli Federico II, Compl. Univ. Monte S. Angelo, 80126 - Napoli, Italy}

\date{Accepted XXX. Received YYY; in original form ZZZ}

\pubyear{2018}

\begin{document}
\label{firstpage}
\pagerange{\pageref{firstpage}--\pageref{lastpage}}
\maketitle

\begin{abstract}
The complexity of the cosmological scenario regarding cosmic strings (CSs) stands still in the way of a complete understanding.
We describe here a promising strategy for the possible detection of these elusive physical entities. It is based on the search of strong gravitational lensing events in the location area of the CS candidate (CSc-1), which was declared in a previous work by CMB analysis. Using photometric and geometric criteria, we identified  pairs of candidates of lensed galaxies (LGCs) in the ``string field'' (SF), which were then compared with the average density of background galaxy pairs in a set of ``control fields'' (CFs). 
We found an excess of $22\%$ (per sq. deg.) of the LGCs in SF, which exceeds the estimated cosmic dispersion. 
We also found that the number of LGCs is in excess of $29.2\%$ in the angular separation bin $[8'', 9'']$.
Finally, we analysed the possibility of a preferred orientation of the line connecting the centres of the LGCs. The orientation is statistically significant for an angular separation bin $[4'',6'']$. Therefore, we found two ``windows'' for the preferred angular separation for LGCs along the possible CS. However, the confirmation of the gravitational lensing origin of our LGCs requires spectroscopic observations which seem to be justified by the present results. We plan to acquire their spectra as well as to continue the study of the spectral and morphological features of the LGCs in the CSc-1 field and to analyse the other CS-candidates using the same strategy. 
\end{abstract}





\section{Introduction}

The search for cosmic strings (CSs) is one of the intriguing problems of modern astronomy, cosmology, and particle physics. CSs are hypothetical one-dimensional objects at cosmological scales (see \citealt{3}; \citealt{sup}) which, while predicted by the theory (\citealt{du}; \citealt{polc}; \citealt{devi}; \citealt{masha}; \citealt{r1}), have not yet been detected. Their ``zoo'' is quite rich \citep{kibble-ayay}. They can be purely topological entities (endless or infinite and closed loops), formed as a result of phase transitions in the vacuum stages of the expansion and cooling of the early universe, or hybrid topological and field configurations (for example, the ``necklace'', a CS with monopoles at its ends and collections of such elements \citep{kibble-ayay}). There is also the possibility of fundamental F-strings and D-strings \citep{dk} of cosmological sizes, which could be generated during high-energy interactions of the extra dimensions in the early universe.

The most important astrophysical characteristic of a CS is the deficit angle $D$, which occurs in the lensing effect on background galaxies, making, as a result, identical pairs of images on both sides of the CS \citep{vil}:
\begin{equation}
D = \frac{8\pi G \mu}{c^{2}}.
\end{equation}\label{q1}

Here $\mu$ is the total CS mass per unit length (linear density) which is proportional to the square root of the CS energy as $( m_{string}/m_{Planck} )^2 = G\mu \ll 1$ (for GUT scale CS the energy will be$~10^{16}$GeV), $G$ is the Newtonian gravitational constant, and $c$ is the speed of the light. Throughout the paper we adopt the Planck units $\hbar = c = 1$. 

In this paper we discuss the search of gravitational lensing events in the field of the best our CS-candidate found in the CMB data \citep{my_best}, hereinafter referred to as CSc-1 (Cosmic String candidate No. 1). 

In gravitational lensing events by CSs, the angular distance between two lensed images is linearly proportional to the deficit angle $D$ multiplied by a combination of the linear distances from the observer to the string-lens $R_q$, \citep{vil, gott, rid} 
\begin{equation}
\Delta \Theta = D \sin \alpha \frac{R_q - R_s}{R_q},
\end{equation}\label{da}
where $\alpha$ is the angle between the vector coinciding with the observer and the center of the background source and the CS.


CS signature can be found by searching for an excess of strong gravitational lensing events: the so-called chain or ``Milky Way of gravitational lenses'' \citep{8}. A lens shall in fact appear every time a CS happens to be placed between the observer and a remote galaxy which forms a small enough angle with the CS \citep{5,6}. The lens chain appears because a CS is expected to possess a cosmological length much wider that the galaxy angular scale (topological CSs of the smaller sizes cannot be studied by methods of gravitational lensing, because they should be located too far away). 

By itself, an excess of lensed galaxy candidates pairs (hereafter referred to as LGCs) is not the proof of the presence of a CS. Clearly, the true gravitational nature of lens candidates must be established by a precise spectroscopic analysis. But even when pairs of objects are found to exhibit identical spectra and the same redshift, we cannot rule out the possibility that this is by chance \citep{4}. The only ``smoking gun'' for a CS is the observation of special cuts in outer isophotes of the lensed image \citep{6}. For this purpose, high angular resolution images of the LGCs are in order.\\

In order to choose the region where to search for an excess of LGCs, we used our previous analysis of the CMB data. A CS could manifest itself as a weak trace in the anisotropy map of CMB radiation (\citealt{71}; \citealt{91}), due to the Kaiser-Stebbins effect of the relic photons which form a step-like distribution (a ``jump'') of the temperature. This is a model-independent property of any CS because all of them lie between the observer and the surface of the last scattering. 

Since the amplitude of the ``jump'' is expected to be very small, two different methods for CS searching in the CMB data exist: 1) the contribution of the CS energy to the total energy of the universe (for details and modern restrictions on the CS tension see \citealt{a-h}; \citealt{7}, and 2) the search for individual CSs.

The Planck group \citep{11}, has investigated the contribution of the CS networks to the full energy of the universe. Their cumulative restriction for the CS angular spectrum is $G \mu \le 1.5 \times 10^{-7}$ for Nambu-Goto and $G \mu \le 11 \times 10^{-7}$ for a semi-local CS; it is instead $G \mu \le 2 \times 10^{-7}$ for Abelian-Higgs model (\cite{che-to}, \cite{ri}). The dynamics of the CS networks from different models has been simulated using an approach derived from the lattice gauge theory \citep{moriarty}. The corresponding Lagrangian density was transformed into a Hamiltonian density and discretized on a
periodic cubic lattice \citep{a-h}. As a result, each model of the CS network is characterized by the function of its own stress-energy tensor from the simulations \citep{ri}. Then, this function is incorporated in the simulations of the CMB anisotropy (for example, with CMBFAST \citep{CMBFAST}), and finally the common simulated CMB spectra are optimized under cosmological parameters (including the CS network parameter) to be consistent with observational data \citep{11}. The CS networks simulations are always model-dependent, badly applied to nontopological CSs, and based on observationally unverifiable initial conditions. It is not in contradiction with our understanding of the early universe the fact that there are no CS networks at all and only few or even just one CS in the whole universe. Finally taking into account the modern observational data (both radio data from WMAP and Planck missions, and the optical data), from our point of view the CS network approach appears not good for CS search.

To describe the second approach a few comments are in order. One of the effective methods for the searching of individual CSs is based on proper convolution techniques. By Haar convolution we have already provided an observational evidence of a semilocal CS with $G\mu \le 7.36 \times 10^{-7}$ \citep{first, my_best} (theoretical justification is in \cite{kibble-ayay}). Using the technique, the CS candidate appears as a continuous line that represents the best match between the Haar step-like function and the ``jump'' on CMB map (the MHF procedure). The modified Haar harmonic is most sensitive to the appearance of discontinuities in radio survey data because an anisotropy induced by a solitary CS represents a sequence of zones of decreased and increased temperature. In order to detect the signal, the power smeared out over all the harmonics must be ``gathered'' to make use of the signal full power. 
For our purposes, the MHF is a realization of the first harmonic of the Haar system of orthogonal functions with cyclic shift. The search for a CS at each point requires multiple convolutions with a rotation of the circle, which corresponds to a shift in the ``jump'' in the Haar function \citep{first}. As a result, we found several CS candidates, which are present in both independent radio sets, WMAP and Planck \citep{my_best}. The best CS candidate CSc-1 extends from $(\alpha= 11 : 29 : 03, \delta = +15 : 23 : 37)$ to $(\alpha = 10 : 57 : 47, \delta = +25 : 03 : 51)$,  with $\delta T/ T \sim 40 \mu K$ (Fig. \ref{fig:CSc-1}). This candidate is the issue of the test being addressed in the optical analysis by the present paper using the gravitational lensing events.

The paper is organized as follows. In Section 2 we provide an analysis of the validity and consistency of the fields from the SDSS data set to be used in the search of gravitational lensing events. In Section 3 we present the detailed optical analysis of closed pairs of similar sources in the field of the CSc-1, using a strategy of statistically comparison of two field sets: (a) the control fields, without any CS candidates, and (b) those where our CSc-1 possibly lies. In the Conclusions we summarize the results and report our arguments which favour the CSc-1 to be a CS. In the Appendix we describe the nonparametric statistical method, the fast rank criterion, which we use in the data analysis.


\section{Optical analysis of gravitational lens candidates in CSc-1 field}

The most promising candidate CSc-1 has been analysed by optical method to look for any excess of gravitational lensing events. The details of the search are described below.

\subsection{Preparation of dataset} 

\begin{figure*}
\centering
\includegraphics[width=13cm]{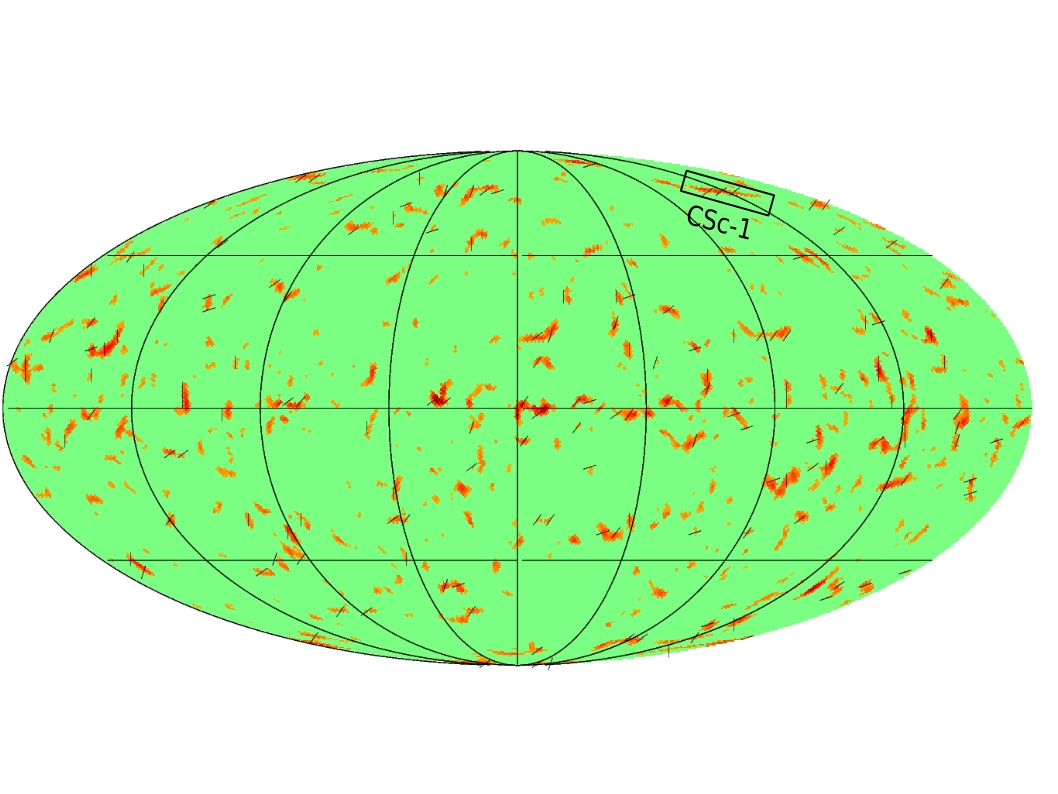}
\caption{The Mollweide projection of the CMB anisotropy Planck map after the MHF convolution. The CS candidate CSc-1 is marked by a rectangular
frame. The amplitude of the CSc-1 is $\sim 40 \mu K$ over an average CMB background
of the order of $100 \mu K$. The short dashes all over the picture indicate the temperature gradients.}
\label{fig:CSc-1}
\end{figure*}

Optical analysis to find sources of strong gravitational  lensing in the region containing the CSc-1 is pursued using a photometric catalogue of galaxies extracted from the Data Release 12 (DR12) of the Sloan Digital Sky Survey (SDSS) which has an angular resolution of $1.3''$
\footnote{http://www.sdss.org/dr12/imaging \label{foo:sdss}}$^{,}$\footnote{http://skyserver.sdss.org/dr12/en/tools/chart/navi.aspx}. We covered an area of $S_{SF}=16.45$ sq. deg., made by 31 overlapping $1^{\circ} \times 1^{\circ}$ sub-fields (assembling the ``string field'', SF) from $(\alpha= 11 : 29 : 03, \delta = +15 : 23 : 37)$ to $(\alpha = 10 : 57 : 47, \delta = +25 : 03 : 51)$, Fig. \ref{fig:CSc-1}.

This field size has been chosen because of the resolution of $1^{\circ}$ used for all sky radio data, but increasing a processing time or processing the selected areas instead of the whole sky, in future studies it will be possible to increase this resolution.\\

Besides the CSc-1 field (SF), we selected 29 ``control fields'' (CFs), approximately $1^{\circ} \times 1^{\circ}$ each, covering a total area of $S_{CF}=30.06$ sq. deg. around and outside the CSc-1 area. The CFs are the ordinary regions of the sky where there is no indication (from our CMB analysis) of the presence of a CSs. These fields will be used to measure the average density of objects in the sky in order to verify the presence of an excess of LGCs. Their centers are reported in Table \ref{cfcentre}.\\

\begin{table}
\begin{center}
\begin{tabular}{c c c}
\hline

$N_{CF}$ & $RA$ & $Dec$ \\ 
$ $ & $[ h:m:s ]$ & $ \left [ ^{\circ}: { }': { }'' \right ]$\\
\hline 
\hline
$	1	$	&	$	10: 56: 16.000	$	&	$	+15: 47: 36.90	$	\\
$	2	$	&	$	11: 06: 50.510	$	&	$	+13: 25: 38.70	$	\\
$	3	$	&	$	11: 43: 50.900	$	&	$	+27: 39: 42.80	$	\\
$	4	$	&	$	12: 36: 26.070	$	&	$	+07: 26: 50.60	$	\\
$	5	$	&	$	12: 43: 16.160	$	&	$	+07: 13: 20.80	$	\\
$	6	$	&	$	16: 48: 00.000	$	&	$	+51: 00: 00.00	$	\\
$	7	$	&	$	14: 12: 00.000	$	&	$	+34: 00: 00.00	$	\\
$	8	$	&	$	00: 36: 00.000	$	&	$	+39: 00: 00.00	$	\\
$	9	$	&	$	08: 00: 00.000	$	&	$	+20: 00: 00.00	$	\\
$	10	$	&	$	10: 22: 20.390	$	&	$	+55: 27: 36.82	$	\\
$	11	$	&	$	10: 46: 52.110	$	&	$	+56: 48: 06.03	$	\\
$	12	$	&	$	12: 20: 13.800	$	&	$	+62: 45: 32.82	$	\\
$	13	$	&	$	13: 13: 08.520	$	&	$	+56: 52: 39.47	$	\\
$	14	$	&	$	14: 41: 14.410	$	&	$	+57: 29: 13.74	$	\\
$	15	$	&	$	14: 55: 23.690	$	&	$	+43: 52: 18.17	$	\\
$	16	$	&	$	16: 05: 57.170	$	&	$	+40: 55: 21.36	$	\\
$	17	$	&	$	08: 48: 58.430	$	&	$	+50: 50: 44.99	$	\\
$	18	$	&	$	16: 43: 19.080	$	&	$	+42: 29: 10.73 	$	\\
$	19	$	&	$	16: 33: 32.860	$	&	$	+32: 08: 41.12	$	\\
$	20	$	&	$	10: 52: 24.800	$	&	$	+54: 37: 02.21	$	\\
$	21	$	&	$	12: 14: 41.820	$	&	$	+60: 32: 56.74	$	\\
$	22	$	&	$	13: 04: 18.970	$	&	$	+55: 01: 47.58	$	\\
$	23	$	&	$	14: 43: 03.700	$	&	$	+43: 53: 05.77	$	\\
$	24	$	&	$	15: 54: 41.140	$	&	$	+42: 02: 56.69	$	\\
$	25	$	&	$	09: 06: 46.280	$	&	$	+49: 04: 41.78	$	\\
$	26	$	&	$	09: 04: 13.830	$	&	$	+53: 39: 49.67	$	\\
$	27	$	&	$	09: 48: 07.200	$	&	$	+61: 50: 36.77	$	\\
$	28	$	&	$	16: 32: 17.970	$	&	$	+43: 57: 46.62	$	\\
$	29	$	&	$	16: 25: 03.380	$	&	$	+33: 52: 05.21	$	\\

\hline
\end{tabular}\caption{Centre coordinates of 29 CFs, in the equatorial coordinate system.}
\label{cfcentre}
\end{center}
\end{table}

In order to build our sample of LGCs for each field (either SF or CFs), we started extracting from SDSS-DR12 all the objects classified as ``galaxy'' in PhotoObjAll, the full photometric catalog for SDSS imaging, requiring that each object has:
\begin{itemize}
\item  a photometric redshift with its error;
\item  $Model \_  mag$ magnitudes in all five bands (u, g, r, i and z) with their errors, as they are appropriate to unresolved objects;
\item a Petrosian radius $petrorad<4.5$ arcsec. This limit allows us to reject objects larger than the deficit angle, i.e. $9''$. 
\end{itemize}

\begin{figure*}
\begin{multicols}{2}
    \includegraphics[width=\linewidth]{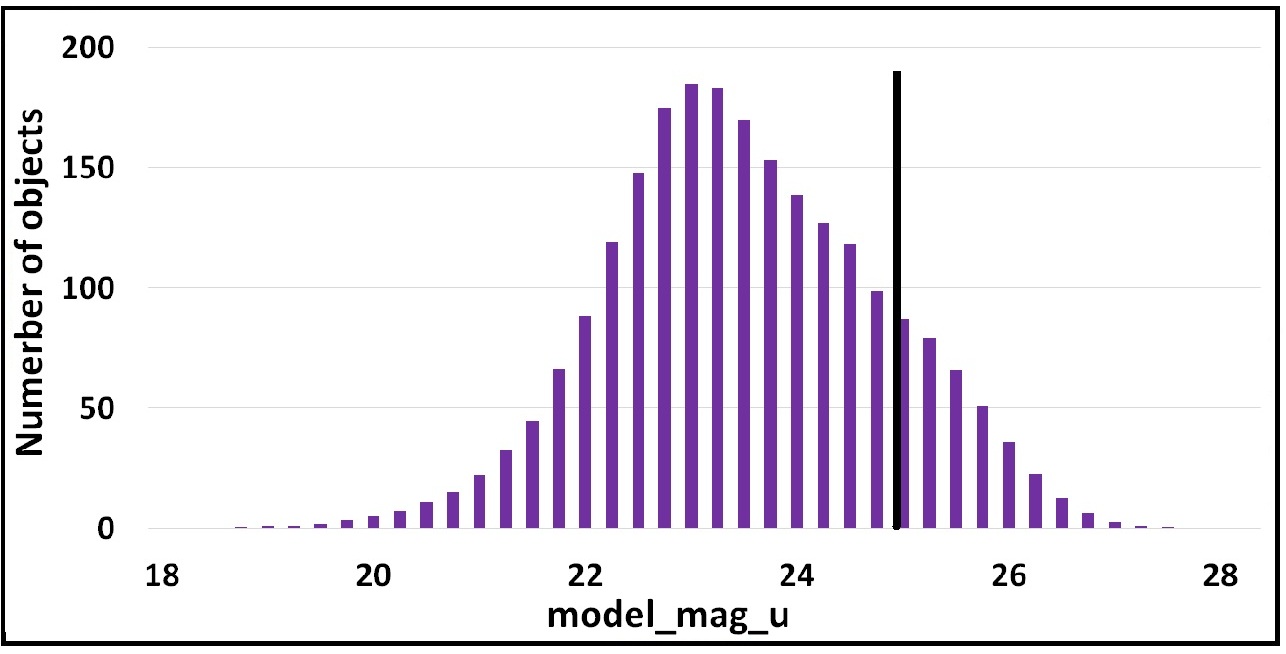}\par 
    \includegraphics[width=\linewidth]{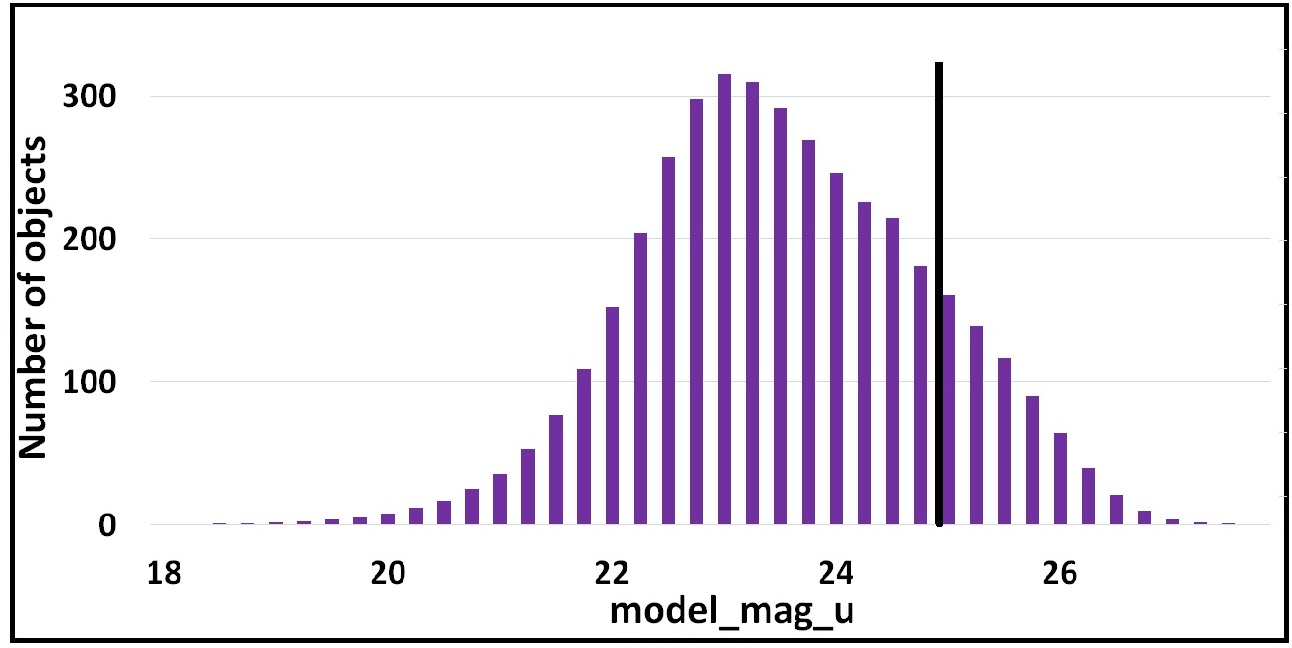}\par 
    \end{multicols}
    
\begin{multicols}{2}
    \includegraphics[width=\linewidth]{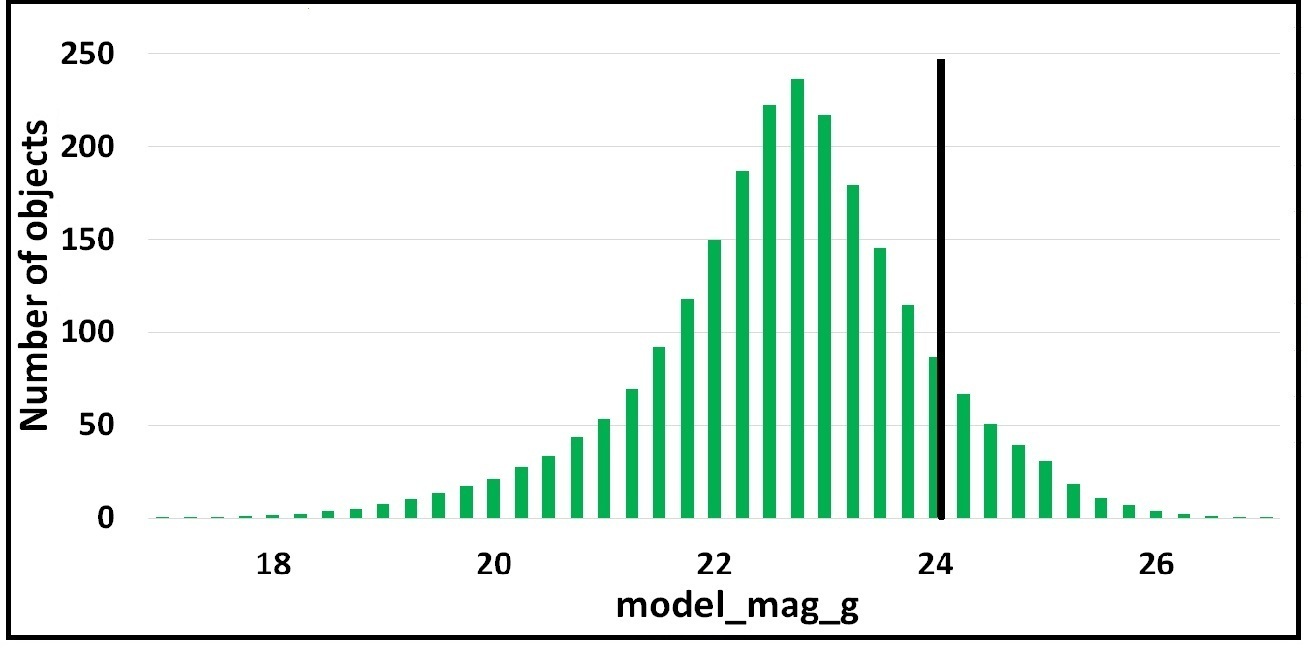}\par
    \includegraphics[width=\linewidth]{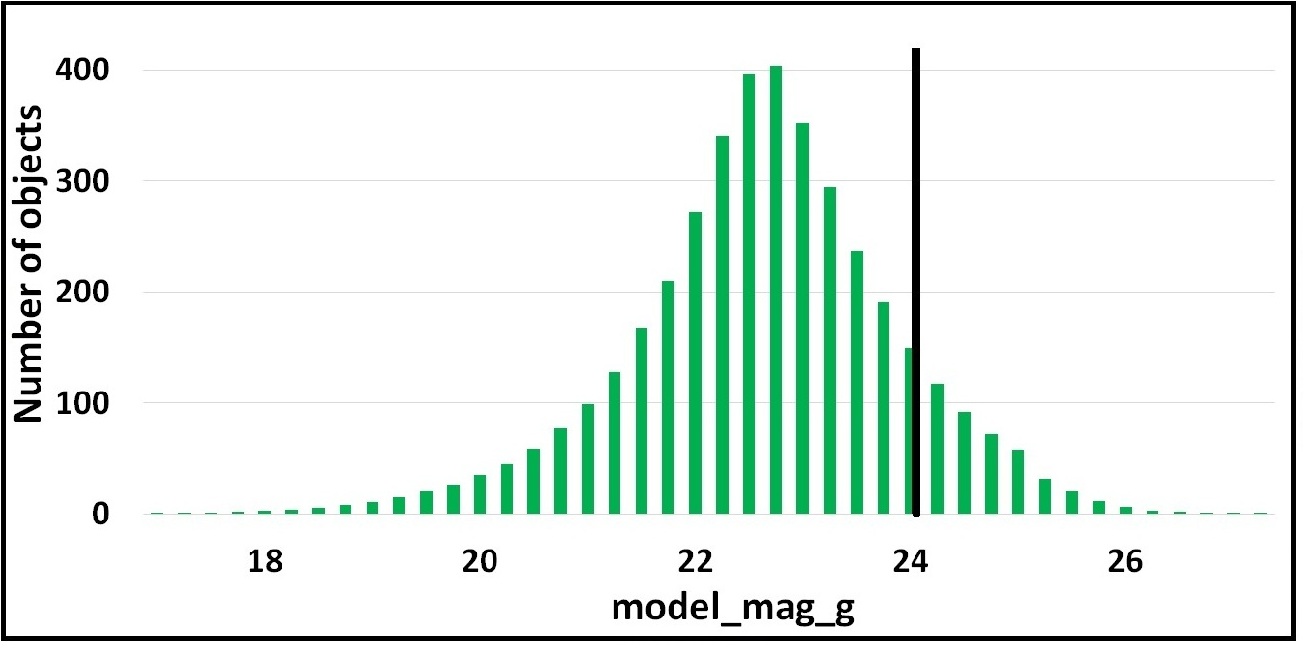}\par
\end{multicols}

\begin{multicols}{2}
    \includegraphics[width=\linewidth]{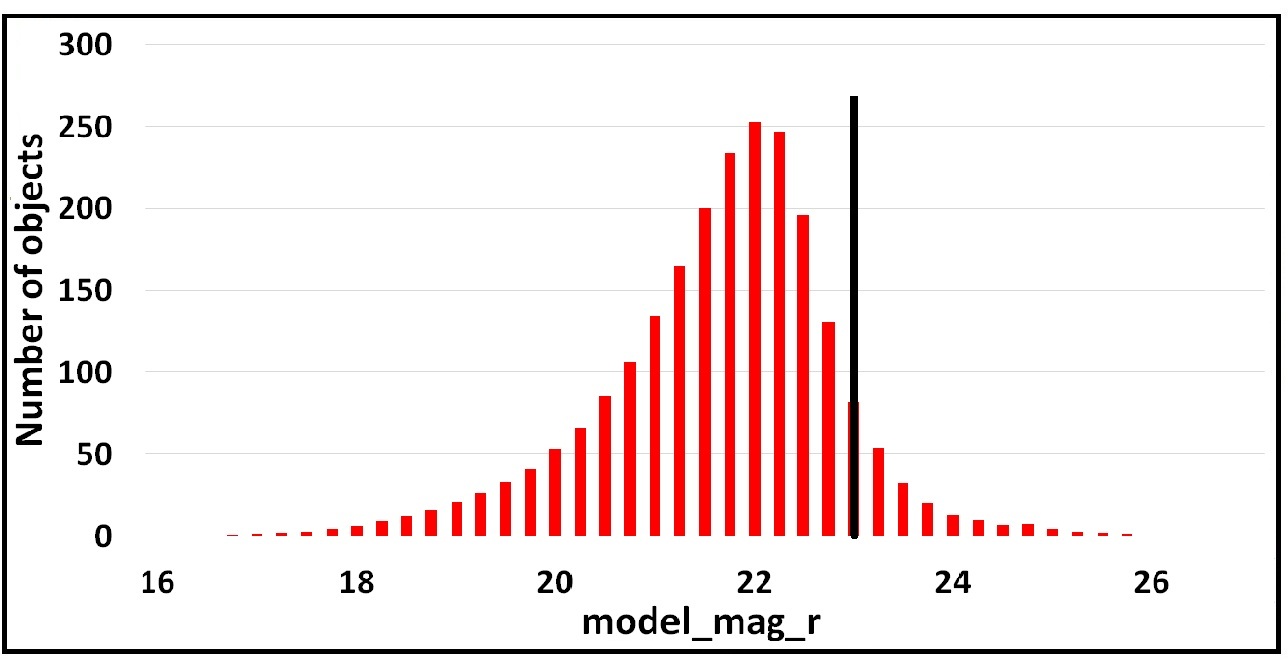}\par
    \includegraphics[width=\linewidth]{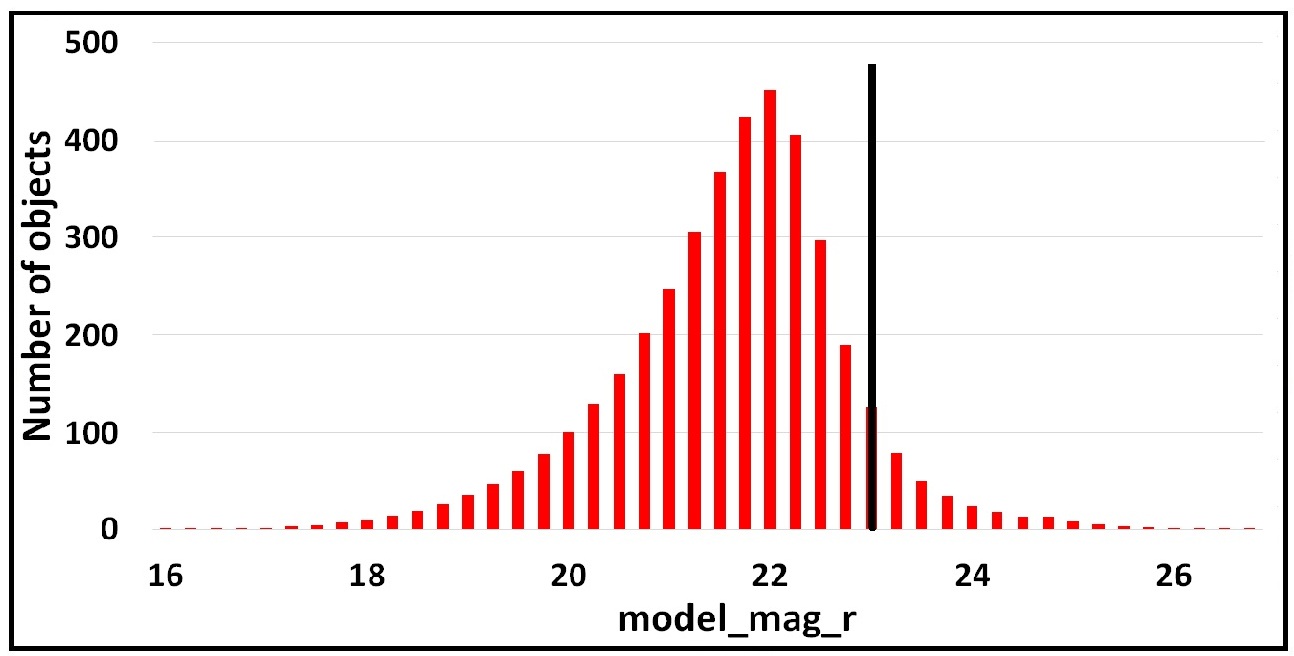}\par
\end{multicols}

\begin{multicols}{2}
    \includegraphics[width=\linewidth]{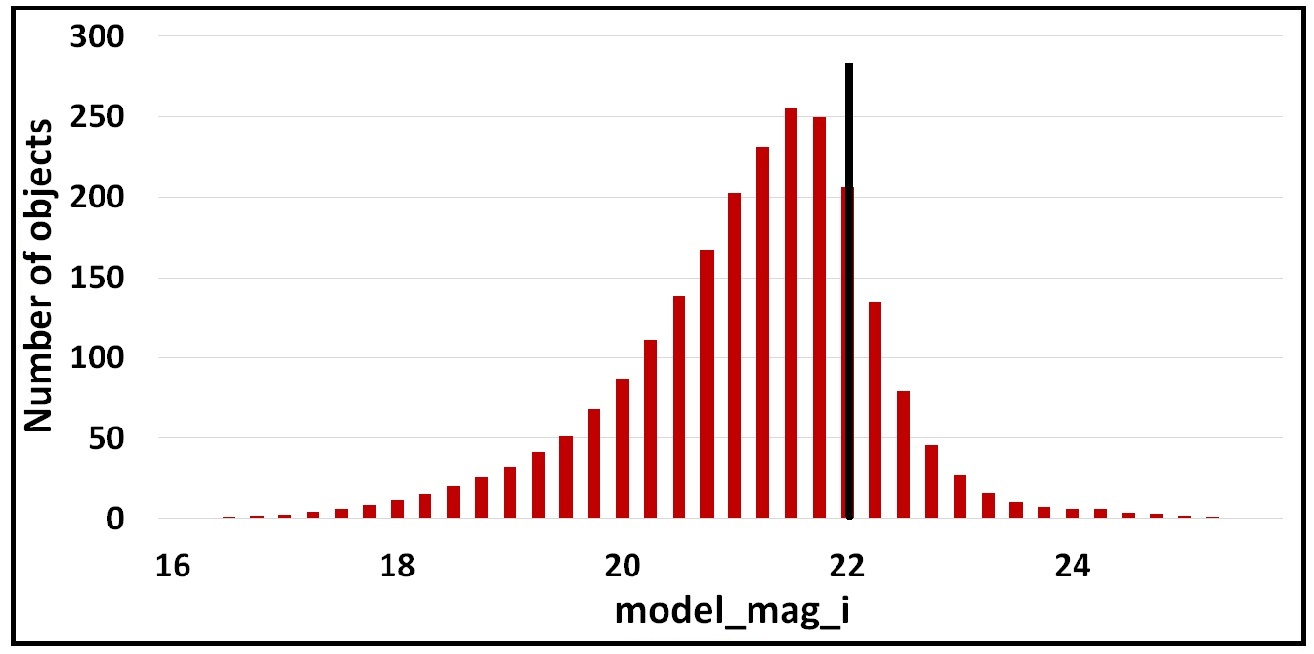}\par
    \includegraphics[width=\linewidth]{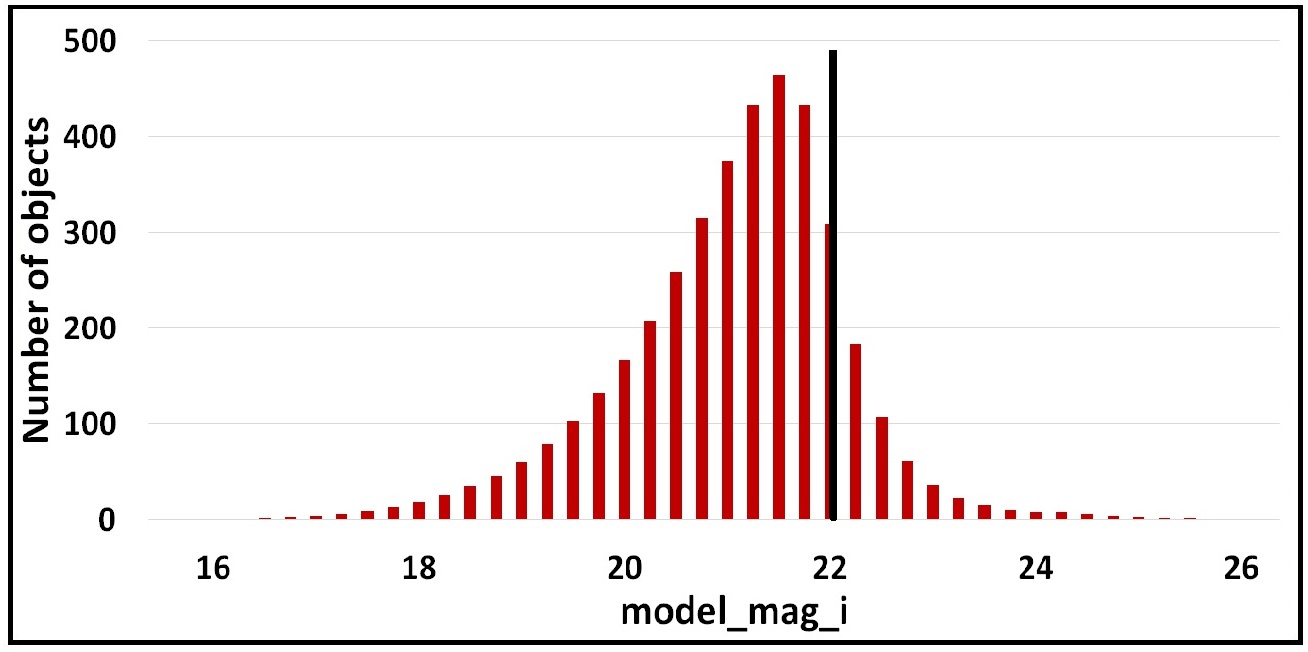}\par
\end{multicols}

\begin{multicols}{2}
    \includegraphics[width=\linewidth]{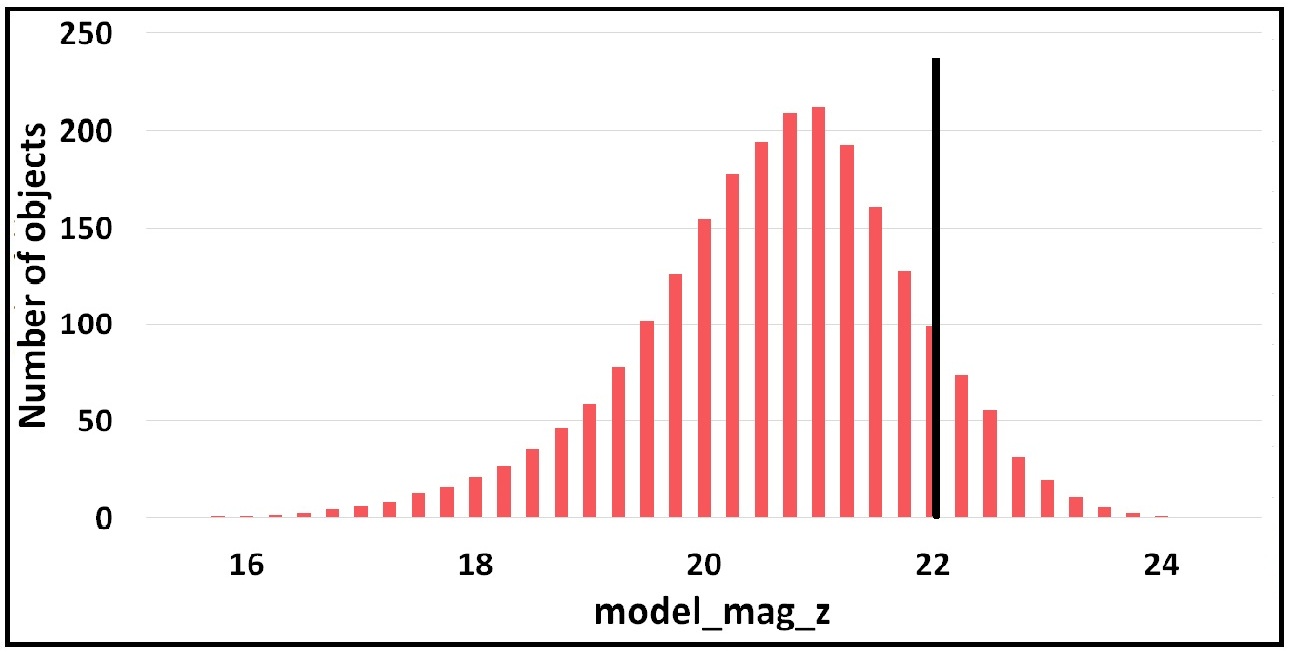}\par
    \includegraphics[width=\linewidth]{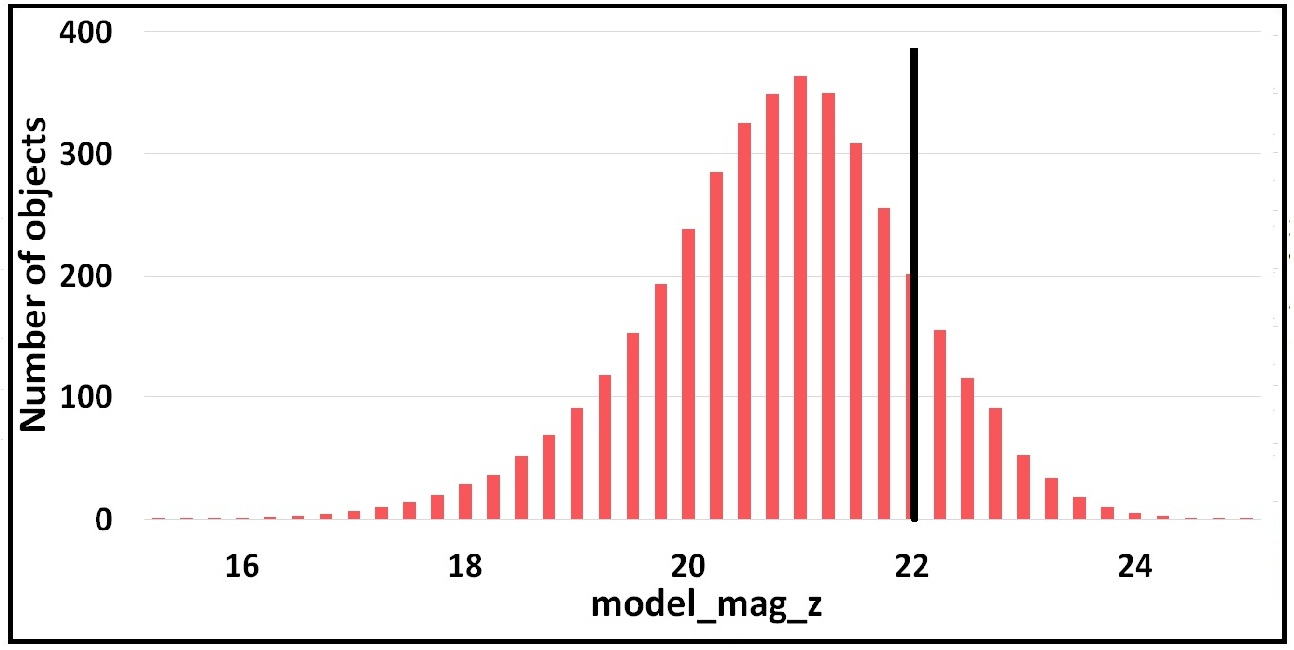}\par
\end{multicols}

\caption{Histograms of $Model \_  mag$ consistency (see text) of the number of galaxies (in unit of $10^{2}$) in the
SF (left panels) and the CFs (right panels) for u, g, r, i, z bands. The black vertical line marks the $Model \_  mag$ limit.}
\label{fig:mag_u_S+lim}
\end{figure*}

Once the double entries due to overlapping sub-fields are removed, the query returned 228,440 objects for the SF and 397,009 objects for the CFs, which give the average densities (per sq. deg.) of 13,887 and 13,207 respectively.
The histograms of Fig. \ref{fig:mag_u_S+lim} plot the number of galaxies as a function of $Model \_  mag$ for each band (u, g, r, i, z) and for both the SF and CF regions. The solid vertical lines shows the $Model \_  mag$ limit which will be discussed in the following.

In order to estimate the reliability of our photometric catalog and the degree of statistical similarity between the distributions in the SF and CF regions, we used the fast rank criterion described in the Appendix. The conclusion is that the hypothesis that the samples are statistically different is rejected with a confidence probability of 90\%.

%
%
%
%

\subsection{Significance test}
The above catalogues may be used to estimate the number of CS strong lensing events that we may expect in the CSc-1 area.
Since we confine our search to objects that we can inspect by eyes, we first set limits to the brightness in each photometric band.
These limits are indicated in Figs. \ref{fig:mag_u_S+lim} as a black vertical line.
Each value corresponds to the minimum magnitude of the LGCs obtained with visual inspection of two control fields, centered at CF$_{1} = (10:56:16, +15:47:35)$ and CF$_{2}  = (11:06:50, +13:25:36)$ respectively, Table \ref{limmag}. We aim to provide a quantitative estimate of the photometric limits and assess the credibility of our data set, through the comparison with the galaxy number counts based on observations obtained with the Sloan Digital Sky Survey by \cite{yas}. In particular, we consider percentages greater than Yasuda's limit of about 17\%, 15\%, 9\%, 6\%, and 6\%, respectively for u, g, r, i, and z band.

\begin{table}
\begin{center}
\begin{tabular}{c c}
\hline

$ Band $ & $Model \_ mag ~ limit$ \\ 
\hline 
\hline
$u$ & $24.6$\\
$g$ & $24.2$ \\
$r$ & $23.0$ \\
$i$ & $22.2$ \\
$z$ & $22.3$\\
\hline
\end{tabular}\caption{Magnitude limits in bands u, g, r, i, z by visual inspection.}
\label{limmag}
\end{center}
\end{table}

\begin{figure}
\includegraphics[width=\columnwidth]{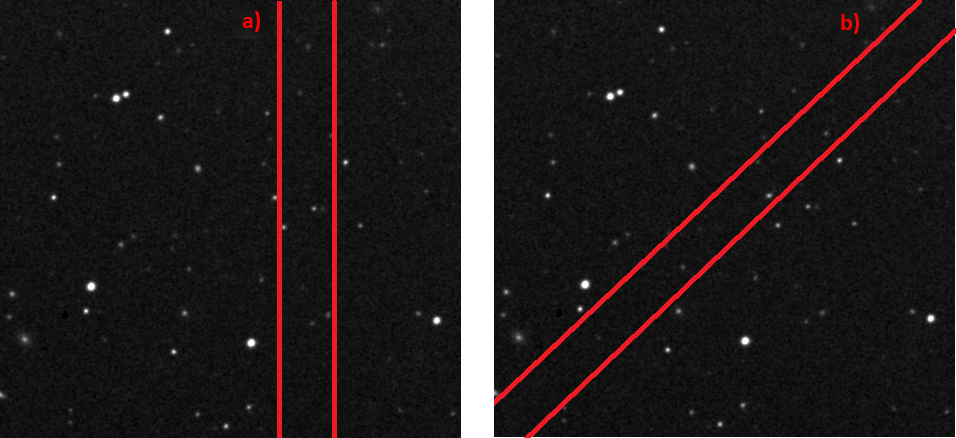}
\caption{Two different cases for a straight CS with a width of $9''$ crossing the field with different directions.}
\label{fig:expect}
\end{figure}

With the above limits, the CFs catalog of 397,009 galaxies reduces to $N_{tot} = 231,381$, corresponding to a galaxy number density of  $<N> = N_{tot}/S_{CF} \approx 7,697 \pm 88$. Since we assume a maximum separation of $9''$ for the LGCs, the number of galaxies in a strip of $1^{\circ} \times 9''$ (mimicking the cosmic string) is  a) $N_a=19.2 \pm 4.4$ or b) $N_b=26.9 \pm 5.2$ depending on the geometrical position of a CS, as showed in Fig. \ref{fig:expect}.


The above values are only lower limits to the expected number of true lensed galaxies since we are assuming that the CS is locally straight and that, according the processed CMB radio maps, the field that we took in account is not affected by the CS. The signal-to-noise ratio for  the expected excess of pairs (always considering a 1 sq. deg.) are $N_{a,b}/\sqrt{N_{a,b}+2\cdot N_{CF}^{sq.deg.}}  \approx 2.4$ in a) case and $\approx 3.4$ in b) case, where $N_{CF}^{sq.deg.}= 21.1$ is the number of galaxy pairs per sq. deg. in the CFs.

This simple test validates our experiment as it demonstrates that, using our data, is possible to detect the presence of a CS at least at $2.4\sigma$ c.l.. Furthermore, the significance test is conducted considering only a 1 sq. deg.. Since CSc-1 subtends a total area of $S_{CF}=30.06$ sq. deg., the overall number of expected lenses increase by a factor $\sqrt{S_{CF}} > 5$.  It is important to stress that the two catalogues of galaxy pairs include either resolved (i.e. extended) objects or pairs that are not as readily recognizable as CS-lensed pairs, due to their faintness. A forthcoming investigation will be devoted to search for discontinuities in the outer isophotes of the resolved LGCs.

\subsection{The procedure}
In this Section, we present the steps of our procedure to detect galaxy pairs (of any kind) in the CFs and SF.
In matching the two catalogs, we adopt the following criteria for the pairs:
\begin{enumerate}
\item their separation must be $2'' < \Delta \Theta < 9''$. The lower limit comes from the resolution of the survey, the upper limit from the relation (\ref{q1}), i.e. we are searching for close objects pairs in according to the tension $10^{-7}<G\mu <10^{-6}$;
\item only objects above the thresholds given above for each band, as reported in Table \ref{limmag}, are retained;
\item the redshifts in each pair must be the same within the errors;
\item also the colours must be the same, within the errors, for the combinations of the SDSS Survey bands (u, g, r, i, z), since gravitational lensing is  achromatic.
\end{enumerate}
The last two conditions translate in:
$$
\left | \left ( m_{x}^{1}-m_{x}^{2} \right )- \left ( m_{y}^{1}-m_{y}^{2} \right ) \right |= \sqrt{(\Delta_{m_x})^2 + (\Delta_{m_y})^2},
$$
with 
$$
\Delta_{m_x} = \sqrt{ (e_x^1)^2 + (e_x^2)^2}, \,\, \Delta_{m_y} = \sqrt{ (e_y^1)^2 + (e_y^2)^2},
$$
where $m_x^1, m_x^2, m_y^1, m_y^2$ are the magnitudes for two galaxies in a pair in the bands $x$ and $y$ respectively, $e_x^1, e_x^2, e_y^1, e_y^2$ are the $1\sigma$ error bars for definition of each magnitude.
Similarily: 
$$
z_{1}-z_{2}= \sqrt{e_{z_{1}}^{2}+e_{z_{2}}^{2}}
$$
where $e_{z_{1}}^{2}$ and $e_{z_{2}}^{2}$ are the redshift errors of the two pair objects.

\begin{figure*}
\centering
\includegraphics[width=\textwidth]{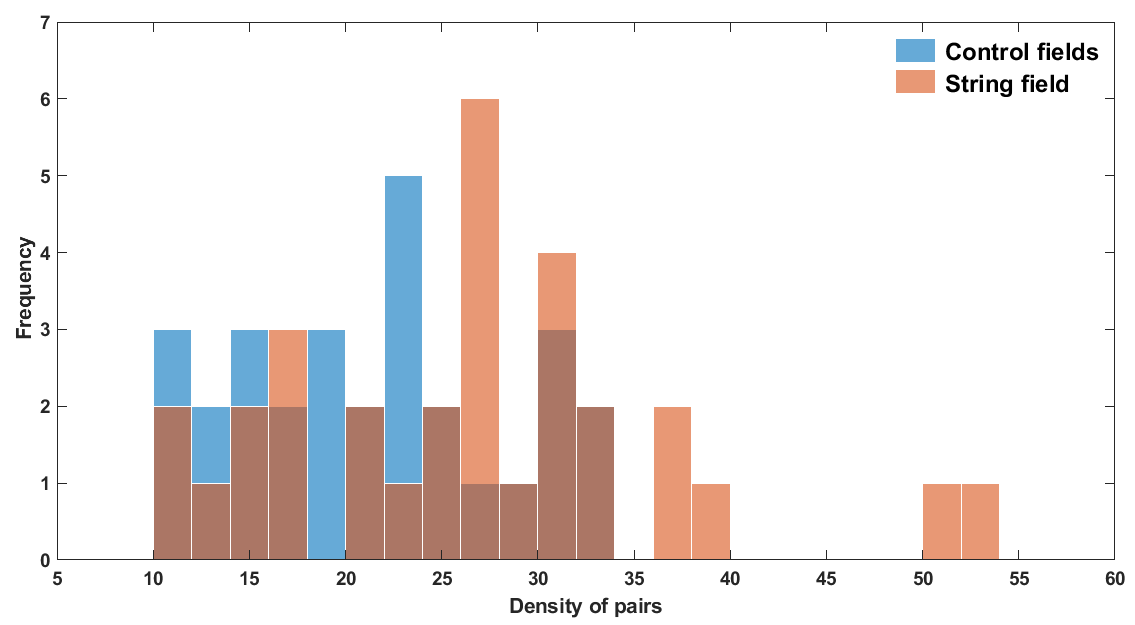}
\caption{Histogram of the number density of LGCs in the SF and in the CFs.}
\label{fig:Density}
\end{figure*}

\begin{figure*}
\centering
\includegraphics[width=\textwidth]{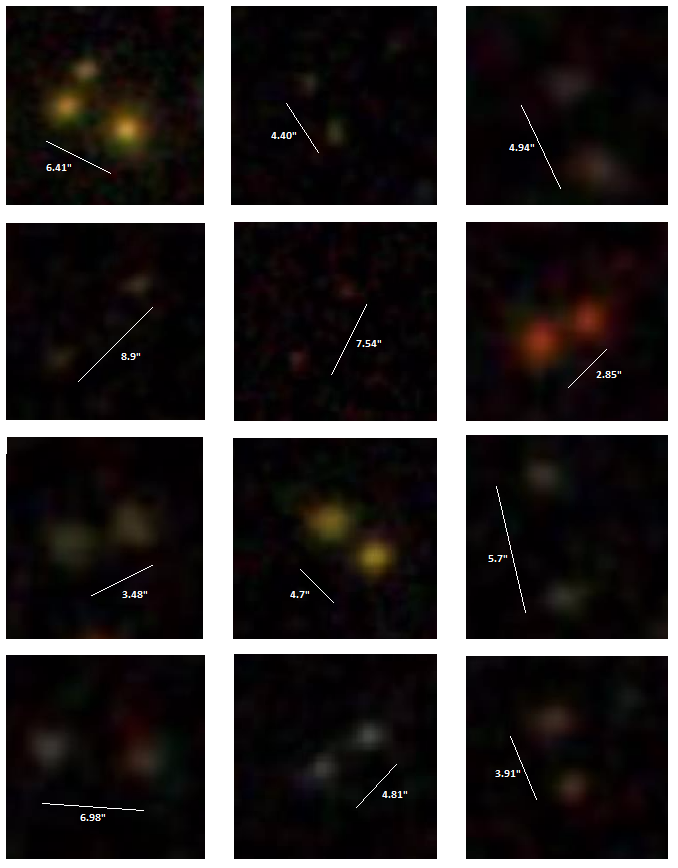}
\caption{Candidate lens galaxies pairs from SDSS DR12. The white dash indicates
the separation between the two components of the pair.}
\label{fig:Coppie}
\end{figure*}

The procedure leads to a list of $N_{SF}= 424$ LGCs in the total SF area, $S_{SF}=16.45$ sq. deg., and $N_{CF}= 635$ LGCs in the total  CF area, $S_{CF}=30.06$ sq. deg.. Fig. \ref{fig:Coppie} shows some examples of LGCs for different deficit angles found via an automatic procedure, checked on the SDSS DR12 Navigate Tool. 
The numbers of LGCs in the SF and in the CFs, normalized to the unit angular surface (1 sq. deg.), provide the following distributions, Fig. \ref{fig:Density}. 


The tail in the distribution of SF pairs is apparent.
The statistical fast rank criterion (see Appendix) shows that the two distributions are significantly different.

The mean number density of LGCs are $\left \langle N_{SF} \right \rangle=26.6\pm1.8$ in SF region and $\left \langle N_{CF} \right \rangle=21.2\pm1.3$ in CF region. 

The number of LGCs per sq. deg. in CSc-1 region (SF) and in background region (CF) are, respectively: $N_{SF}^{sq.deg.}= 25.8$ and  $N_{CF}^{sq.deg.}= 21.1$. Thus, there is a statistically significant excess of $22\%$ (per sq. deg.) of LGCs in SF with respect to the number of LGCs in the CFs, corresponding to an excess of 4.6 pairs per sq. deg. with a significance level $ \sim 3\sigma$.
This statistical result is in agreement with the expected value estimated above.

\subsection{Analysis of the separation between the galaxy pair objects}
\begin{figure}
\includegraphics[width=\columnwidth]{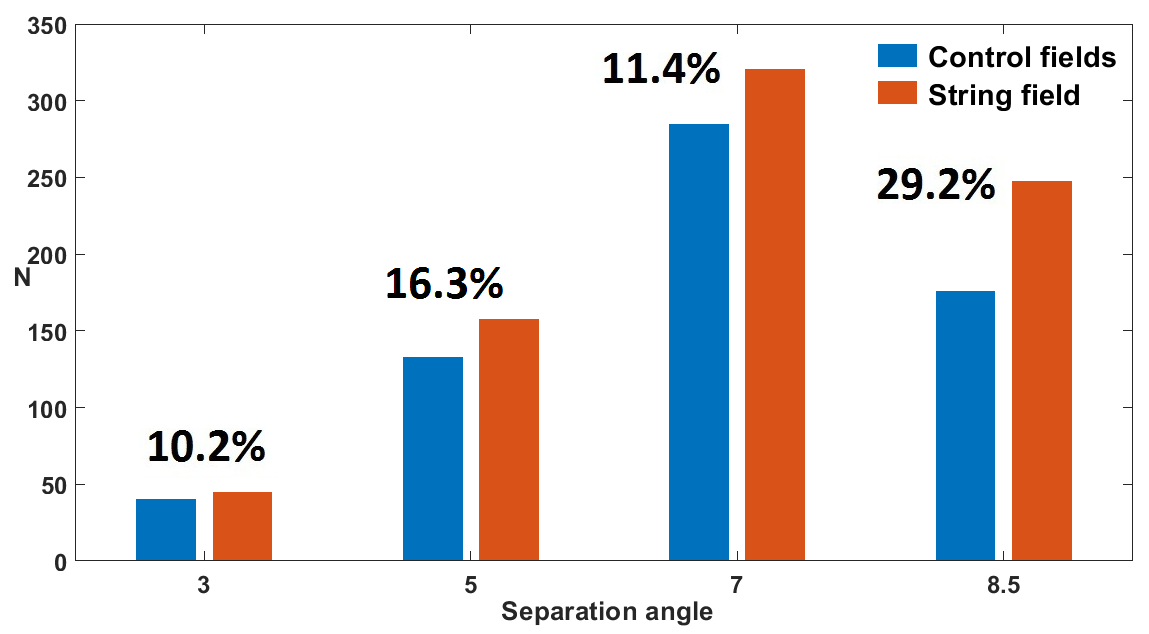}
\caption{Histogram for different separations (in arcseconds) of the pairs within SF
and CFs. The largest excess is for the separation $[8'', 9'']$.}
\label{fig:Separazioni_s_t}
\end{figure}

We want to investigate here if there is a preferred range of separations between the components of the pairs. To this end, we compare the distributions
 for CFs and SF, rescaled to the same area. We find (see Fig. \ref{fig:Separazioni_s_t}) 
 an excess of $29.2\%$ in the CSc-1 region in the deficit angle range $[8'',9'']$.  It should be a first indication of the physical properties of CSc-1 as a CS candidate. There is also a second possible ``window'' for CS deficit angle $[4'',6'']$ if we assume that the CS is straight, with an excess of $16.3\%$.


\subsection{Analysis of orientation of galaxies pairs}
\begin{figure}
\includegraphics[width=\columnwidth]{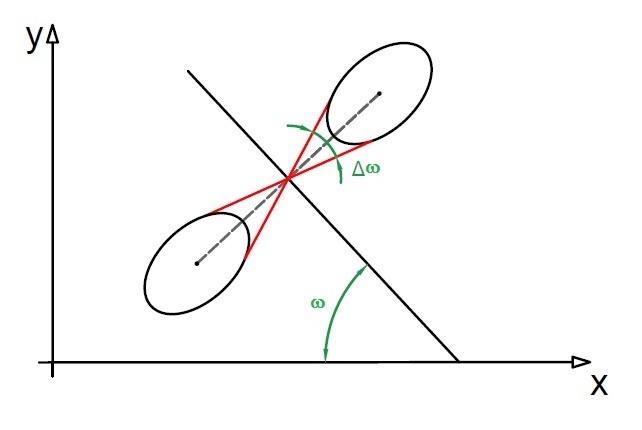}
\caption{Sketch of a pair of galaxies, LGC, forming an angle $\omega$ with the horizon, together with its error $\Delta \omega$. }\label{fig:im_coppia}
\end{figure}

Under the assumption that the CS is straight, we can find its direction by checking whether there is a preferred orientation between close pairs of galaxies along straight lines across the sky. For each field, we calculate the relative number of LGCs with the same angle of inclination $\omega$ as a function of PSF, Fig. \ref{fig:im_coppia}.

For a line $l$ which connects the centres of the pair components (with the specified coordinates $\left \{ x_{1}, y_{1} \right \}$ and $\left \{ x_{2}, y_{2} \right \}$ for the $1^{st}$ and $2^{nd}$ component, respectively), the equation of the perpendicular line $l_{p}$ is:
$$
y=-\frac{\left ( x_{2}-x_{1} \right )}{y_{2}-y_{1}}\cdot x+\frac{\left ( x_{2}-x_{1} \right )}{y_{2}-y_{1}}\cdot x_{1}+y_{1}
$$
The accuracy $ \Delta \omega $ on the slope of the line $l_{p}$ depends on the the resolution of the image (PSF of the nearest star) by the equation:
$$
\left ( \frac{FWHM}{\Delta \Theta } \right )=tg\left ( \frac{\Delta \omega }{2} \right ),
$$
where $ \Delta \Theta  $ is the angular distance between the components of the pair and FWHM is the angular resolution. In spherical coordinates:
$$
\Delta \Theta =\sqrt{\left ( y_{1}-y_{2} \right )^{2}+cos^{2}y_{1}\cdot \left ( x_{1}-x_{2} \right )^{2}}.
$$

The inclination angle $ \omega_{i} $ (in arcsec) for each pair $i$, and its error are given by:
$$
\omega_{i}=-arctan\left( \dfrac{x_{i2}-x_{i1}}{y_{i2}-y_{i1}}\right) \pm \Delta \omega_{i},
$$
where
\begin{equation}
\Delta \omega_{i} = 2\arctan\left ( \frac{PSF_{i}}{\sqrt{\left ( y_{i1}-y_{i2} \right )^{2}+\cos ^{2}y_{i1}\cdot \left ( x_{i1}-x_{i2} \right )^{2}}} \right ).
\label{deltabeta}
\end{equation}
In order to estimate the number of pairs which have compatible orientation, we calculate for all pairs the corresponding angle $\omega_i$ and its error $\Delta \omega_i$. 
We report the case for $\Delta \omega_i =0$. 

According to the theory, we expect a random uniform distribution of $ \omega_{i} $ in fields without the CS and a multi-modal distribution or a distribution with a single peak in the fields with a CS candidate. The reason of such a ``multi-modality'' depends on the curvature of the CS. 
In the simplest case of a straight CS, the angles of the lines $l$ should be obviously the same for all the LGCs generated by a CS. 
In a more realistic case of a curved CS, however, one expects that the number of LGCs with certain angles of inclination should overcome the number of LGCs with other angles.

\begin{figure*}
\begin{multicols}{2}
    \includegraphics[width=\linewidth]{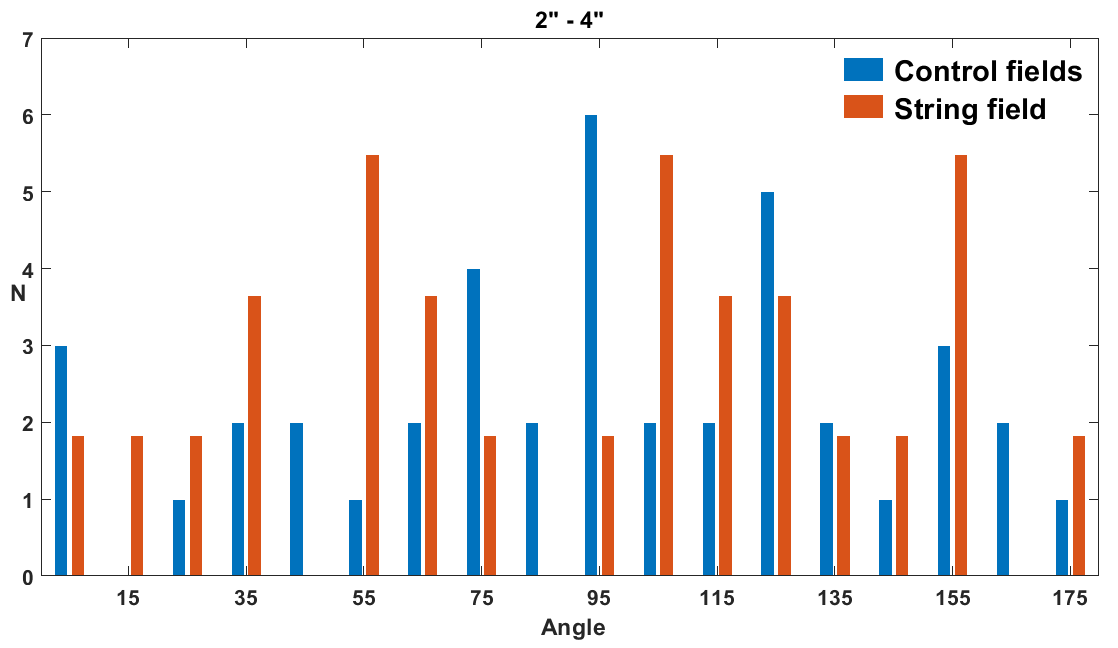}\par 
    \includegraphics[width=\linewidth]{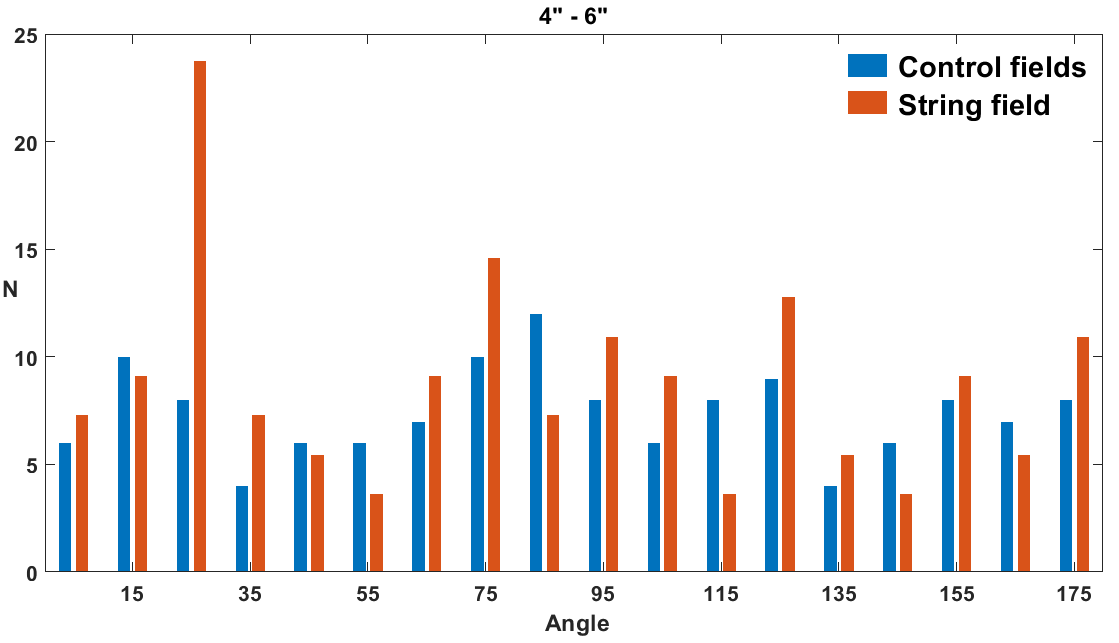}\par 
    \end{multicols}
    
\begin{multicols}{2}
    \includegraphics[width=\linewidth]{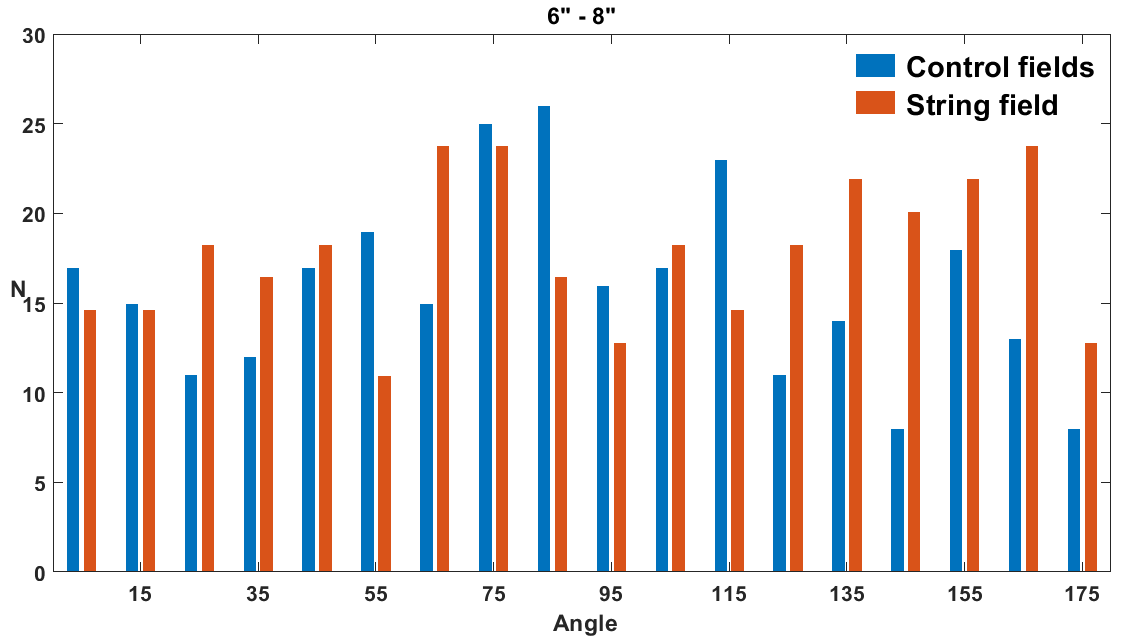}\par 
    \includegraphics[width=\linewidth]{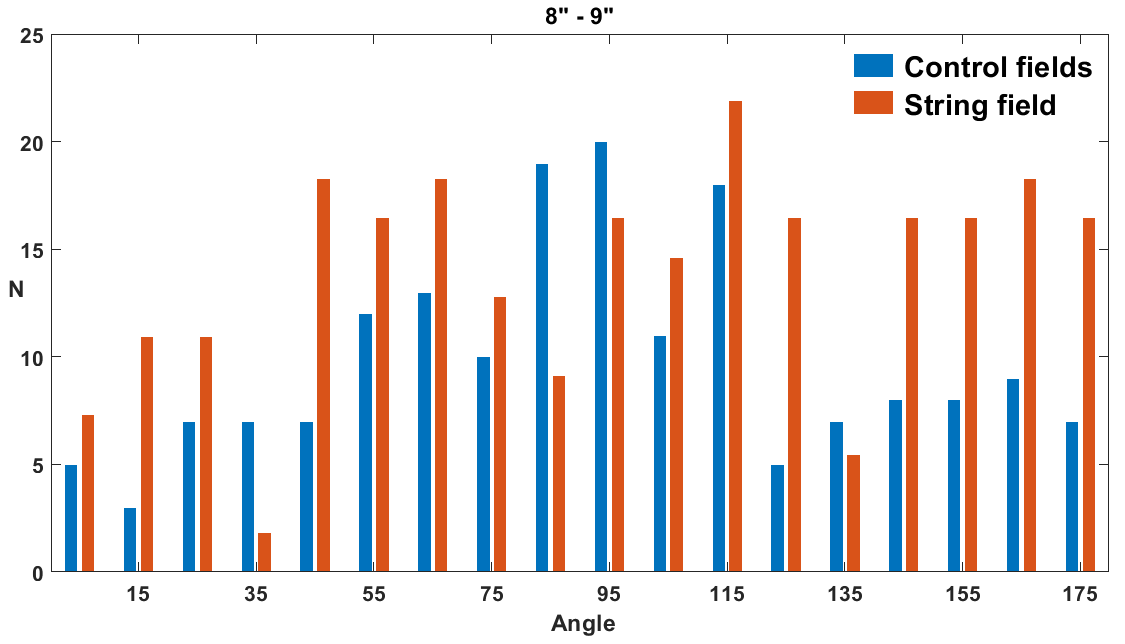}\par 
    \end{multicols}    
    
\caption{Histograms for the number $N$ of the LGCs as a function of the position angle for different separation bins quoted at the top of each panel for the SF and the CFs with $\Delta \omega_i=0$. The angle bin is in degrees.}
\label{fig:2_4} 
\end{figure*}

In the search procedure of the pairs, we considered a rather wide ``window'' for possible distances between the pair components, from $2''$ up to $9''$, dividing the interval into four subintervals: $[2'', 4'']$, $[4'', 6'']$, $[6'', 8'']$, $[8'', 9'']$. For each of them we plot an histogram of the numbers of pairs (LGCs) $N$ vs. the inclination angle $\omega_i$ of the line $l$ (binned in  $10^{\circ}$ intervals: $\omega_i \in \{[170^{\circ},180^{\circ}], [160^{\circ},170^{\circ}], ... [0^{\circ},10^{\circ}]\}$. We report the histograms for $\omega_i$ with $\Delta \omega_i=0$ for the four separation ranges, see Fig. \ref{fig:2_4}. 

    
    
%

We apply a t-test to check the presence of an excess in the distribution of LGCs numbers for a certain inclination angle interval (after normalizing to the same area and checking with the $\chi^2$-test the normal distribution of the samples). The results indicate the statistically significant abundance of $[4'',6'']$ distanced pairs in the interval $\omega_i \in [20^{\circ}, 30^{\circ}]$ in equatorial system, that  in galactic coordinates corresponds to $\omega_i \in [10^{\circ}, 21^{\circ}]$. The calculated statistics for the maximal sample value $n_{max}=23.8$ in the bin $[4'',6'']$ is $t^{*}=3.07>t_{0.05,17}$ where $t_{0.05,17}=2.11$ is the tabular t-quantile for p-value 5\% and for 17 degrees of freedom. Although the excess of pairs in the SF with respect to the CFs pairs in that case is $16.3 \%$, we cannot accept this result without further check. To establish the reliable correlation with the presence of a CS, optical observations with higher precision are needed anyway.


Thus, we obtained two preferred values for the angular separation, for each of them there is a statistically significant excess of candidates for gravitational-lens pairs. In other words we found two possible ``windows'' for the CS deficit angle. 
The positions of the pairs for the corresponding angular distances are shown in the Fig. \ref{fig:Separazioni_s_t}.
At least one of them could be ``wrong''. But without further spectroscopic follow up, it is impossible to determine exactly which angular distance value should be excluded.

Knowing the magnitude of the anisotropy induced by the CSc-1 \citep{my_best}, we can give an estimation of its velocity. In the simplified model, the geometrical combination in equation (\ref{da}) could be taken equal to 1 \citep{4,6}. So the expression (\ref{q1}) defines the angular distance between lensed images. Using the simplest kind of relation between CS anisotropy $\delta T$, CS tension $\mu$, and projection $\beta$ of the CS velocity on the plane perpendicular to the line of sight, we get:
\begin{equation}
\frac{\delta T}{T}=8\pi G\mu \frac{\beta}{\sqrt{1-\beta^{2}}}. 
\label{magnit}
\end{equation}
Here $T=2.73 K$ is the CMB temperature and $\delta T$ has the order of $\mu K$. For the CSc-1 $\delta T/T \approx 40 \mu K$, \citep{my_best}. Thus for $\Delta \Theta =5''$ the CS velocity $\beta$ is $0.5c$ and for $\Delta \Theta = 8.5''$ is $0.3c$.

\section{Conclusions}
CSs are expected to produce well defined and unambiguous observational features. In the CMB, CSs produce well defined step-like discontinuities in the temperature. In fact, the structure of such temperature fluctuations depends on the CS parameters since the fluctuations are affected by the position of the CS with respect to the observer, by the CS velocity and orientation, and by the CS linear density. In optical surveys, the signature of the presence of a CS is a chain of lensed galaxies and sharp edges in well resolved galaxy images.

In this work, we addressed the problem of recognition of the CSs signatures, starting from radio data to optical one. In a previous paper, the convolution of the radio maps, obtained from WMAP and Planck CMB data, with the modified Haar functions, provided a list of preliminary CS candidates with the amplitude $\delta T/T \le 40 \mu K$. For the best one of them, CSc-1, covering the area of $S_{SF}=16.45$ sq. deg., here we carried out the optical analysis, based on the search of sources of strong gravitational lensing on the SDSS-DR12 galaxy catalog. Using a simple algorithm, we identified LGCs that satisfy photometric and geometrical criteria and obtained a list of $N_{SF}=424$ pairs in the total area of string candidate. 
In order to test whether there is an excess of LGCs in SF with respect to the cosmic variance pertaining to non-CS areas, we applied the same search criteria to a set of control fields covering an area of $S_{CF}=30.06$ sq. deg. where we obtained $N_{CF}=635$ pairs. As a result, we firstly checked with these numbers the significance of our experiment, which turns to be better than $2.4 \sigma$ c.l..

We then found that the number density of lens candidates LGCs in the string field exceeds by $22 \%$ that in the comparison fields, in good agreement with the CS scenario. Assuming pure Poissonian fluctuations, the excess of $\sim 5$ pairs per sq. deg. in the CSc-1 region with respect to the control one is significant at $\sim 3\sigma$ level.

The next step was to identify a preferred range, if any, in the separation angle between the components of the each pair, comparing the distribution for CFs and SF. The result was two possible ``windows'' of $29.2\%$ for $[8'', 9'']$ and of $16.3 \%$ for $[4'', 6'']$ angular separations respectively. Using the mean value of these separation ranges, we compute also the CS velocity. By the simplified assumption that the CS is straight, in order to ascertain the presence of an excess of close galaxy pairs in SF for a certain inclination angle range, we plotted, separately for each separation interval, the number of LGCs with an inclination angle $\omega$. Then, we have applied the Student's t-statistics, obtaining a statistically significant excess for the  $[4'', 6'']$ distanced pairs in the interval $\omega \in [20^{\circ}, 30^{\circ}]$. 

Even if no definitive conclusion can be drawn at this point, this study provides intriguing hints on the fact that CSc-1 might indeed be a cosmic string.
A stronger test of the true gravitational nature of the lens candidates requires a spectroscopic investigation. 
But the {\it experimentum crucis} remains the detection of cuts in the outer isophotes of found resolved lens candidates. For this purpose, high angular resolution images of the lensed sources are in order.


\section*{Acknowledgements}

The authors acknowledge the support from the Program of development of M.V. Lomonosov Moscow State University (Leading Scientific School 'Physics of stars, relativistic objects and galaxies').

We also would like to thank the Prof. Maurizio Paolillo (Department  of  Physics,  University  of  Naples  Federico  II) and Arina Morgunova (Sternberg Astronomical Institute, M.V. Lomonosov Moscow State University) for  the useful discussion.

\section*{Appendix}

To compare samples with unknown distribution laws, people usually use nonparametric statistical methods, testing hypothesis of equality in the position and  scale parameters of the corresponding samples (\citealt{quen}; \citealt{si}). We used here the fast rank criterion, which is based on the analysis of the sequence of ranks, $R_i$, of sample values and not on the sample values themselves. The elements of the two samples, i.e. $n$ and $m$ are combined in a single sample and then, in order to test the hypothesis of no displacement of the mean, they are ranked in ascending order: $x_1 \le x_2 \le x_i \le ... \le x_{n+m}$ with $i = R_i$.
Instead, to test the hypothesis of no scale distortion of two samples, the combined sample is ranked in ascending order: $x_1 \le x_2 \le x_i \le ... \le x_{n+m}$ and then, it is rewritten as $x_1, x_n, x_{n-1}, x_2, x_3, x_{n+m-2}, x_{n+m-3}, x_4, x_5, ...$.

In both cases, the statistic $d^{*}=|d/s_d|$ is the standard normally distributed quantity ($d^{*}\sim N(0,1)$), where $s_d=\sqrt{(\sum R_{(1)} + \sum R_{(2)}) \cdot(1/n+1/m)/6}$,  $d= \sum R_{(1)}/n - \sum R_{(2)}/m$, $\sum R_{(1)}$ and $\sum R_{(2)}$ are the rank sum for the first and second samples respectively in the combined samples. 

If the calculated value $d^{*}>u_{1-p/2}$, where $u_{1-p/2}$ is the tabular Gaussian quantile for the given p-value, then the hypothesis of ``displacement of the mean'' (the hypothesis of ``scale distortion'', respectively) of two samples is not rejected.  

The effectiveness of the fast rank criterion is more than 86\% for any non-Gaussian distributions and is 95\% for Gaussian distributions.

We applied this criterion to compare the galaxy distributions for each band (u, g, r, i, z) for the SF and CF regions, respectively. Here, we report the case of the galaxy distributions in u band (the worst case): the calculated statistic is $d^{*}=1.60$ for the hypothesis of ``displacement of the mean'' and $d^{*}=1.53$ for the hypothesis of ``scale distortion''. For $p=0.1$, the tabular Gaussian quantile is $u_{1-p/2}=1.64$ which is greater than both calculated statistics. Hence, the hypothesis that the samples are statistically different is rejected with a confidence probability 90\%.

We also used this criterion to compare the densities in the two distributions of the LGCs in the SF and in the CFs. The calculated statistics is $d^{*} = 2.41$ for the hypothesis of ``displacement of the mean''. For $p=0.018$ the critical value of tabular Gaussian quantile is $u_{1-p/2}=2.37$. Hence, the hypothesis that the samples are statistically different is not rejected with a confidence probability 98.2\% (and even more so, it is not rejected for the previous $p=0.1$).

The results of the complete statistical analysis with several independent non-parametric criteria are presented in the Table \ref{criterion}. We also analyzed the consistency of these criteria. For this purpose, we simulated several statistically identical samples and compared them. The results proved the consistency of the criteria used; see Table \ref{criterion1}.

\begin{table*}
\begin{center}
\begin{tabular}{c c c c c c}
\hline
 Method/ & No          & Large number  & Small number & Critical  &  Method/   \\ 
  criterion  &   grouping  & of grouping   & of grouping  & value     &  criterion     \\ 
             &             & intervals     & intervals    & /region   &  effectiveness \\
\hline 
\hline
Fast rank             & $|-2.1818|$ & $2.2747$ & $|-2.9892|$   & $1.96$ & $0.86$ \\
\hline
Mann-Whitney          & $|-2.1819|$ & $3.2451$ & $14$          & $1.96$ for big samples,     & $0.95$ \\
   -Wilcoxon          &             &          &               & $[18,63]$ for small samples$^*$ &  \\                               
\hline
Iman                  & $|-2.2191|$ & $3.5246$ & not valid for & $1.995$                     & $0.95$\\
approximation         &             &          & small samples &                             &  \\
\hline
Van der Waerden       & $|-2.1168|$ & $2.3840$ & $3.0554$      & $1.96$ for big samples      & t-statistics \\
                      &             &          &               &                             & effectiveness \\
                      &             &          &               &                             & for big samples \\
\hline
\end{tabular}\caption{Statistical results of the comparison of two samples: the numbers of LGCs in the SF and in the CFs. We used four independent non-parametric statistical methods (column 1) and different ways of splitting into bins (in the table we present three typical cases: the number of bins is equal to the number of elements of the initial samples (column 2), a large number of bins (column 3) and a small number of bins (column 4)). The column 5 is the tabular critical value/interval. The column 6 is the method effectiveness. In all cases, the observational statistics exceeds the tabular values, which means that the two distributions are different. The only borderline case is the value of the Van der Waerden criterion; however, its effectiveness for small samples has not been proven. $^*$For small samples instead of the Mann-Whitney-Wilcoxon, we use the Mann-Whitney U-statistics, for which it is defined the critical interval.}
\label{criterion}
\end{center}
\end{table*}

\begin{table}
\begin{center}
\begin{tabular}{c c c c c}
\hline
 Method/ &  1  & 2  & 3 & Critical    \\ 
  criterion  &     &    &   & value \\ 
\hline 
\hline
Fast rank             & $0.064$ & $0.2704$ & $0.3660$ & $1.96$  \\
\hline
Mann-Whitney          & $0.045$ & $0.0350$ & $0.0134$ & $1.96$  \\
   -Wilcoxon          &         &          &          &   \\                               
\hline
Iman                  & $0.0316$ & $0.0348$ & $0.0133$ & $1.995$ \\
approximation         &          &          &          &  \\
\hline
Van der Waerden       & $1.2960$ & $0.325$ & $0.414$   & $1.96$ \\
\hline
\end{tabular}\caption{Statistical results of comparison of two simulated statistically identical samples. We present three examples of the simulations (1, 2,3) for four independent non-parametric statistical methods (for the large number of grouping interval, see Table \ref{criterion} caption). In all cases, the observational statistics is less the tabular values, which means the consistency of the criteria.}
\label{criterion1}
\end{center}
\end{table}

\newpage

\end{document}